\documentclass[12pt]{article}
\usepackage[dvips]{graphicx}
\begin{document}
\title{\bf Light element synthesis in baryon isocurvature models}
\author{D. Lohiya \& P. Kumar
\footnote{I.U.C.A.A., Pune}\\ 
DAMTP, University of Cambridge, Cambridge CB3 0WA \\
email: D.Lohiya@damtp.cam.ac.uk	}
\date{}

\maketitle      
\begin{center}
Abstract\\
\end{center} 

The prejudice against baryon isocurvature models is primarily because of
their inconsistency with early universe light element nucleosynthesis 
results. We propose that incipient low metallicity (Pop II) star
forming regions can be expected to have environments 
conducive to Deuterium production by spallation, up to levels observed 
in the universe. 

\section{Introduction:}

Early universe (standard big-bang) nucleosynthesis [SBBN] is widely
regarded as a major success for the 
Standard Big Bang [SBB] Model. As presented, SBBN results look 
rather good indeed. The observed light element abundances are 
taken to severely constrain cosmological and particle physics 
parameters. Deuterium, in particular, is regarded as
the ideal ``baryometer'' for determining the baryon content of the 
universe \cite{olive}. 
This follows from the fact that deuterium is burned away whenever
it is cycled through stars, and from a belief, that there are no 
astrophysical sites (other than SBBN), capable of producing it 
in its observed abundance \cite{eps}. 

Primary Isocurvature Baryon 
models [PIB] are characterised by a baryon density
that saturates dynamic mass bounds from galactic clusters. These models
can provide concordance with CMB anisotropy \cite{HUTHESIS} and offer
a plausible account for structure formation in the universe \cite{ppbles}. 
The model does away with with the need for non - baryonic dark matter. 
One could consider a PIB model with $\Omega_b \approx 0.2$ to $0.3$ 
and comply with the theoretical prejudice in favour of a flat model by 
adding a cosmological constant $\Omega_\Lambda = 1 - \Omega_b$.  
The problem with such a model is that the high baryon content gives low yields 
for Deuterium in a primordial nucleosynthesis scenario. 

There have been suggestions that an early generation of massive black holes may
form just after recombination in the PIB models. These could be sites on which
accreting gas would emit radiation that could generate $D$ and $^3He$ by 
photo disintegration of $^4He$ \cite{gned}. The purpose of this article it to 
show that the environment of incipient Pop II stars are sufficient to produce 
these elements by spallation reactions. One thus does not need to
invoke black holes for this purpose at all.

Confidence in SBBN stems primarily 
from $D$, $^7Li$ and $^4He$ measurements.  $D$ abundance is 
measured  in solar wind, in interstellar clouds  and, more 
recently, in the inter-galactic medium \cite{geiss,hogan}. 
The belief  that no realistic astrophysical process other than 
the Big Bang  can  produce sufficient $D$ lends 
support to its primordial origin. Further, 
$^7Li$ measurement [$^7Li/H \sim 10^{-10}$] in Pop II stars 
\cite{spite}  and the consensus \cite{yang} over the primordial 
value for the $^4He$ ratio $Y_p \ge 23.4 \%$ 
(by mass) suggest that
light element abundances are consistent with SBBN over nine 
orders of magnitude. This is  achieved by adjusting just
one parameter,  the baryon entropy ratio $\eta$. Alternative 
mechanisms for $^7Li$ production that are 
accompanied by a co-production of $^6Li$ with a later depletion 
of $^7Li$ have fallen out of favour. The debate on depletion
of $^7Li$ has been put to rest by the observation of $^6Li$ in a 
Pop II star \cite{smith}. Any depletion of $^7Li$ would have to 
be accompanied by a  complete destruction of the much more 
fragile $^6Li$. Within the SBBN scenario therefore, one seeks to 
account for the abundances of $^4He$, $D$, $^3He$ and $^7Li$ 
cosmologically, while $Be$, $B$ and $^6Li$ are generated  by 
spallation processes \cite{olive1}.

These results do meet with occasional scepticism
[see eg. \cite{primak} for problems with BBN]. 
Observation of $^6Li$, for example, requires unreasonable 
suppression of astrophysical destruction of $^7Li$. On the 
other hand, the production of $^6Li$ would be accompanied by a 
simultaneous production of $^7Li$ comparable to observed levels 
\cite{rees}. This raises doubts about using observed $^7Li$ 
levels as a benchmark to evaluate SBBN. 

Further the best value of $^4He$ mass fraction, 
statistically averaged and extrapolated
to zero heavy element abundances, 
hovers around $.216 \pm .006$ for Pop II objects \cite{rana}.
Such low $^4He$ levels have also been reported     
in several metal poor HII galaxies \cite{pagel}. For example 
for  SBS 0335-052 the reported value is $Y_p = 0.21 \pm 0.01$ 
\cite{ter}. Such small values for $^4He$ would not
lead to any concordant value for $\eta$ consistent with bounds on
$^7Li$ and $D$. Of course, one could still explore a 
multi-parameter non-minimal  SBBN instead of the minimal model 
that just uses $\eta$ for a single parameter fit. Non-vanishing 
neutrino chemical potentials have been proposed, on the one hand, 
to be ``natural'' 
parameters for such a venture. Adherents of minimal SBBN have criticised
these conclusions in \cite{yang,ter} on grounds of reliance on statistical 
over-emphasis on a few metal-poor objects with a high enough 
$^4He$ abundance. On the other hand, there 
are objects reported with abysmally low $^4He$ levels. This is 
alarming for  minimal SBBN. For example, levels of $^4He$ 
inferred for $\mu$ Cassiopeia A \cite{ter} and from the emission
lines of several quasars \cite{peim} are as low as 5\% and 
10 - 15\% respectively. Such low levels would most definitely 
rule out SBBN. At present one excludes such objects from SBBN 
considerations on grounds of ``our lack of understanding''of the 
local environments of these objects. As a matter
of fact,  one has to resort to specially contrived explanations to
account for low $Y_p$ values in quasars. Considering that a host 
of mechanisms for light element synthesis are discarded on grounds
of requirement of special ``unnatural'' circumstances \cite{eps}, 
it does not augur to have to resort to special explanations to 
contend with low $^4He$ emission spectra. This comment ought to 
be considered in the light of much emphasis that is 
laid on emission lines from nebulae with low metal content 
\cite{yang}. Quasars most certainly qualify for such candidates. 
Instead, one merely seems to  concentrate on classes of Pop II 
objects and HII galaxies
that would oblige SBBN. Until dependence of light 
element abundance on sample and statistics is dispensed with and 
/ or more fully understood, we should not close our eyes to 
alternative solutions.

We end this overview of the status of SBBN  with a few comments.
Firstly the low metallicity that one sees in type II stars and 
interstellar clouds poses a problem in SBBN. There is
no object in the universe that has low abundance 
[metallicity] of heavier elements as is produced 
in SBB. One relies on some kind of re-processing, much later in 
the history of the universe, to get the low observed  
metallicity in,  for example, old clusters and 
inter-stellar clouds. This reprocessing could take place in a generation 
of very short-lived type III stars conceived for this purpose. 
Such a generation of stars
may also be necessary to ionise the intergalactic medium. 
The extrapolation of $^4He$ 
abundance in type II objects and low metal (HII) galaxies, 
to its zero heavy metal abundance 
limit, presupposes that reprocessing and production of heavy
elements in type III stars is not accompanied by a significant 
change in the $^4He$ levels. A violation of this assumption,
i.e. a minute increase in $^4He$ during reprocessing (even as low
as 1 - 2 \%) would rule out the minimal SBBN. As a matter of fact, 
it is possible to account for the entire pre-galactic $^4He$ 
by such objects \cite{wagoner}.

Finally, of late \cite{lem}, the need for a careful scrutiny and 
a possible revision of the status of SBBN has also been suggested 
from  the reported high abundance of $D$ in several $Ly_\alpha$ 
systems. It may be difficult to accommodate such high abundances
within the minimal SBBN. Though the status of these observations 
is still a matter of debate and (assuming their confirmation), 
attempts to reconcile the cosmological abundance of deuterium 
and the number of 
neutrino generations within the framework of SBB are still on, 
a reconsideration of alternate routes to deuterium 
as described in this article
could well be worth the effort. This is specially in consideration
of the stranglehold that Deuterium has on SBB in constraining the 
baryon density upper limit to not more than some 3 to 4 \% . This
constraint has been used in SBB to make out a strong case for non -
baryonic dark matter to close the dynamic mass estimates at galactic and
cluster scales. Making out a case for CDM from observed (local 
environment sensitive) estimates of Deuterium runs the risk of
{\it ``building
a colossus on a few feet of clay''\cite{borner}}.

%\section{Getting rid of the Deuterium Stranglehold:}
\section{{\bf Deuterium Production}}

For a high baryon density model like the PIB model to clear the observed 
Deuterium constraints, the 
desired amount of Deuterium would have to be produced much later 
in the history of the universe. To this effect, we recall spallation 
mechanisms that were explored in the 
pre - 1976 days \cite{eps}. Deuterium can indeed be produced by 
the following spallation reactions:
$$
p + ^4He \longrightarrow D + ^3He; ~~ 2p \longrightarrow D + \pi^+;
$$
$$
2p \longrightarrow 2p + \pi^o,~ \pi^o \longrightarrow 2\gamma,~
\gamma +^4He \longrightarrow 2D.
$$There is no problem in producing Deuterium all the way 
to observed levels. The trouble is that under most conditions 
there is a concomitant over - production of $Li$ nuclei and 
$\gamma$ rays to unacceptable levels. Any later destruction of 
lithium in turn completely destroys $D$. As described in 
\cite{eps}, figure (1) exhibits relative production of $^7Li$  
and $D$ by spallation. It is apparent that the production of 
these nuclei to observed levels, and without a collateral 
gamma ray flux, is possible only if the incident (cosmic ray or any
other) beam is energised to an almost mono energetic value of 
around 400 MeV. A model that requires nearly mono energetic 
particles would be rightly considered $ad~hoc$ and would be hard to 
physically justify. This presents a ``no-go'' result. There is no 
way to produce $D$ without overproducing $Li$.

However, lithium production occurs by spallation of protons over 
heavy nuclei as well as spallation of helium over helium:
$$
p,\alpha ~+ ~C,N,O \longrightarrow Li~+~X;~~ 
p,\alpha ~+ ~Mg,Si,Fe \longrightarrow Li~+~X;~~
$$
$$
2\alpha \longrightarrow ^7Li ~+~p; ~~ \alpha ~+~D \longrightarrow p ~+~^6Li;
$$
$$
^7Be + \gamma \longrightarrow p + ^6Li; ~~ ^9Be + p \longrightarrow
\alpha +^6Li.
$$
Essentially, the ``no-go'' argument of Epstein et al \cite{eps} used 
$Y_\alpha /Y_p \approx .07$ ($\approx 28\%$ by weight)
in both the incident particle
flux as well as the ambient medium, besides the canonical solar
heavy element mass fraction in the target cloud.
The absence or deficiency of heavy nuclei in a target cloud and 
deficiency of alpha particles in the incident beam would clearly 
suppress lithium production. Such conditions could well have 
existed in the environments of incipient Pop II stars. 

Essential aspects of evolution of a collapsing cloud to form a low
mass Pop II star is believed to be fairly well understood 
\cite{feig,hart}. The formation
and early evolution of such stars can be discussed in terms of
gravitational and hydrodynamical processes. A protostar would 
emerge from the collapse of a molecular cloud core and would be 
surrounded by high angular momentum material forming a 
circumstellar accretion disk with bipolar outflows.
Such a star contracts slowly while magnetic fields play a 
very important role in regulating collapse of the accretion disk 
and transferring the disk orbital angular motion to collimated 
outflows. A substantial fraction of the accreting matter is 
ejected out to contribute to the inter - stellar medium.

Empirical studies of star forming regions over the last twenty 
years have now provided direct and ample evidence for MeV 
particles produced within protostellar and T Tauri systems 
\cite{Terekhov,Torsti}. The source of such accelerated 
particle beaming is understood to be violent magnetohydrodynamic 
(MHD) reconnection events. These are analogous to solar magnetic
flaring but elevated by factors of $10^1$ to $10^6$ above levels 
seen on the contemporary sun besides being up to some 100 times 
more frequent. Accounting for characteristics in the meteoritic 
record of solar nebula from integrated effects of particle 
irradiation of the incipient sun's flaring has assumed the status 
of an industry. Protons are the primary component of particles 
beaming out from the sun in gradual flares while $^4He$ levels are 
suppressed by factors of ten in rapid flares to factors of a 
hundred in gradual flares\cite{Terekhov,Torsti}. Models of young 
sun visualises it as a much larger protostar with a cooler 
surface temperature and with a very highly elevated level of 
magnetic activity in comparison to the contemporary sun. It is 
reasonable to suppose that magnetic reconnection events would 
lead to abundant release of MeV nuclei and strong shocks that 
propagate into the circumstellar matter. Considerable evidence 
for such processes in the early solar nebula has been found in 
the meteoric record. It would be fair to say that the 
hydrodynamical paradigms for understanding the earliest stages of 
stellar evolution are still not complete. However, it seems 
reasonable to conjecture that several features of collapse of a 
central core and its subsequent growth from accreting material 
would hold for low metallicity Pop II stars. Strong magnetic 
fields may well provide for a link between a central star, its 
circumstellar envelope and the accreting disk. Acceleration of 
jets of charged particles from the surface of such stars would 
have suppressed levels of $^4He$. Such a suppression 
could be naturally expected if the particles are picked up by 
magnetic reconnection events from 
an environment cool enough to suppress ionised $^4He$ in 
comparison to ionised hydrogen. Ionised helium to hydrogen number 
ratio in a cool sunspot temperature can be 
calculated  by the Saha's ionization formula and the 
ionization energies of helium and hydrogen. This could be as low as
$\approx~ exp(-40)$ and increases rapidly with temperature. Any 
electrodynamic process that accelerates charged particles from 
such a cool environment would yield a beam deficient in alpha 
particles. With $^4He$ content in an accelerated particle beam 
suppressed in the incident beam and with the incipient cloud 
forming a Pop II star having low metallicity in the 
target, the ``no - go'' concern of (Epstein et.al. \cite{eps}) 
is effectively circumvented. As stated before, the ``no-go'' result
used $Y_\alpha /Y_p \approx .07$ in both the energetic incident particle
flux as well as in the ambient medium besides the canonical solar 
heavy element abundance in the target cloud. Incipient Pop II environments may 
typically have heavy element fraction suppressed by more than 
five orders of magnitude while, as described above, magnetic 
field acceleration could accelerate beams of particles deficient 
in $^4He$.

One can thus have a broad energy band - all the way from a few 
MeV up to some 500 MeV per nucleon as described in the Figure (1), 
in which acceptable levels of deuterium could be ``naturally'' 
produced. The higher energy end of the band may also
not be an impediment. There are several astrophysical processes 
associated with gamma ray bursts that could produce $D$ at high 
beam energies with the surplus gamma ray flux a natural by product.

\section{{\bf Conclusions:}}

Our understanding of star formation has considerably evolved 
since 1976. SBBN constraints need to be reconsidered in view of 
empirical evidence from young star forming regions. These models 
clearly imply that spallation mechanism can lead to viable and 
natural production of Deuterium and Lithium in the incipient 
environment of Pop II stars. One can consider a baryon dominated PIB
cosmological model in which early universe nucleosynthesis produces the 
desired primordial levels of $^4He$ but very little $D$. 
In such a universe, in principle, spallation mechanism can lead to
Deuterium and Lithium  synthesis up to acceptable levels 
in the environment of incipient Pop II stars.

In SBB, hardly any metallicity is produced in the very early
universe. Metal enrichment is supposed to be facilitated by a
generation of Pop III stars. Pop III star formation from a
pristine material is not well understood till date in spite of a
lot of effort that has been expanded to that effect recently
\cite{sneider}. It is believed that with metallicity below a
critical transition metallicity
$(Z_{cr} \approx 10^{-4} Z_\odot$), masses of Pop III stars would 
be biased towards very high masses. Metal content higher than 
$Z_{cr}$ facilitates cooling and a formation of lower mass Pop II 
stars. In SBB, the route to Deuterium by spallation discussed in 
this article would have to follow a low metal contamination by a 
generation of Pop III stars.

\eject
\begin{center}
\begin{figure}
\resizebox{.8\columnwidth}{!}
{\includegraphics{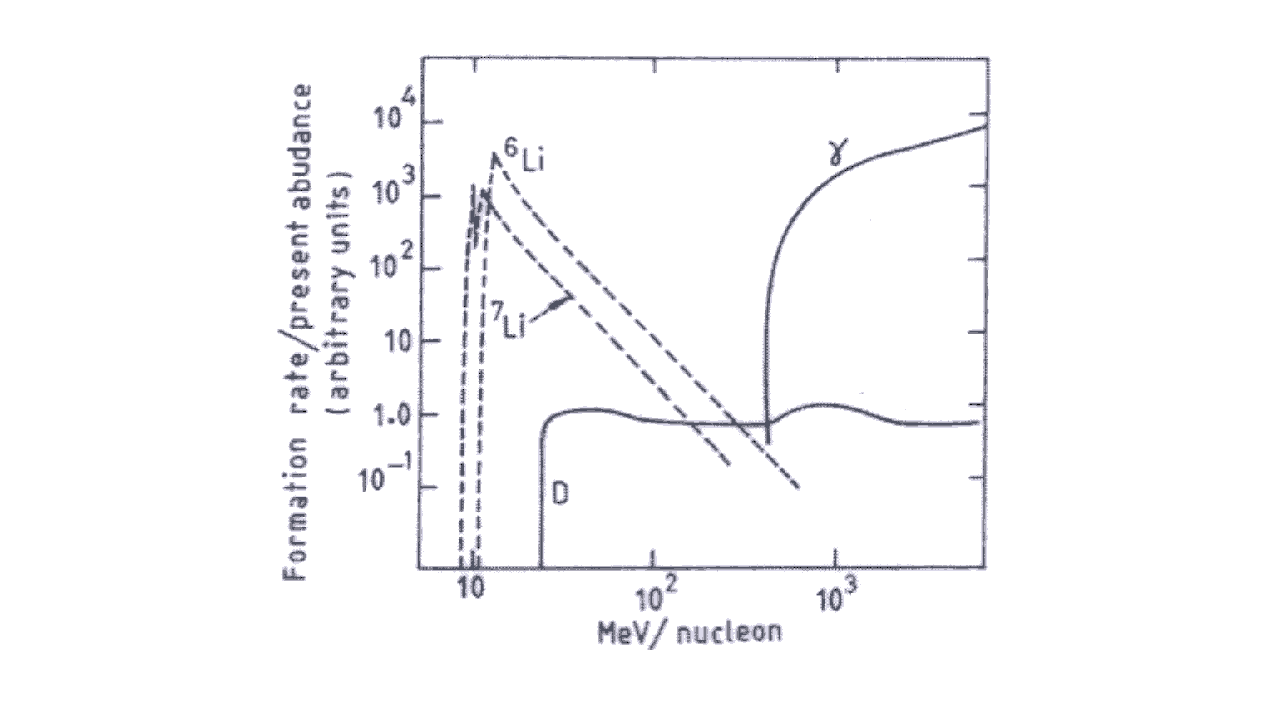}}\\
\title{Figure(1): The ratio of formation of abundances to their 
present values as a 
function of the energy per nucleon of the incident particle. Below $\approx 
400$ MeV, the observed levels of $D$ are accompanied by an over production of 
Li isotopes while slightly above this energy a large gamma ray flux is 
produced} \cite{eps}\\
\end{figure}
\end{center}
\vskip 1cm

\vfil\eject
\section{{\bf Acknowledgment.}} The foundation Fellowship from Clare Hall, 
Cambridge is gratefully acknowledged. 
\vspace{0.5cm}

\bibliography{plain}

\end{document}